\documentclass[11pt,twoside]{article}

\usepackage{asp2006}

\begin{document}
\title{An Eclipsing ``Blue Straggler'' V228 from 47 Tuc: Evolutionary Consideration}   
\author{Marek J. Sarna}   
\affil{N. Copernicus Astronomical Center, Polish Academy of Sciences, Bartycka Str. 18, 00-716 Warsaw, Poland}    

\begin{abstract} 
We perform evolutionary calculations of binary stars to find
progenitors of system with parameters similar to the eclipsing
binary system V228. We show that a V228 binary system may be
formed starting with an initial binary system which has a low main
sequence star as an accretor. We also show that the best fitting
model implies loss of about 50 per cent of initial total orbital
momentum but only 5 per cent of initial total mass.
\end{abstract}


\section{Introduction}   

Blue straggler (BS) stars are defined by their location on the
color--magnitude diagram. These star lie above the main--sequence
turnoff region, a region where, if the BS's had been normaly
single stars, they shuold alredy have evolved away from the main
sequence. V228 is eclipsing binary system -- member of the
globular cluster 47 Tuc. On the color--magnitude diagram of the
cluster, the variable occupies a position near the top of the BS
region. However, an analysis of Ka\l u\.zny et al. (2006) shows
that V228 is a semi--detached Algol--type binary with the less
massive component filling its Rochel lobe. We know several system
of such type (RY Aqr, S Cnc, R CMa, AS Eri).

\section{What do we know about V228 and 47 Tuc?}

Ka\l u\.zny et al. (2007) obtain the following parameters for
primary and seconadry respectively:

\noindent {\bf Primary:} $M_1 = 1.512\pm0.022$; $R_1 = 1.357\pm
0.019$; $L_1/L_\odot = 7.02\pm 0.05 $,

\noindent {\bf Secondary:} $M_2 = 0.20\pm0.007$; $R_2 = 1.238\pm
0.013$; $L_1/L_\odot = 1.57\pm 0.09 $,

\noindent {\bf Orbital period:} 1.150686 days.

For globular cluster 47 Tuc we know that age of the cluster is
between 10 and 14 Gyrs (Gratton et al. 2003, VandenBerg et al.
2006, Ka\l u\.zny et al. 2007), metalicity from Alves--Brito et
al. (2005) is [Fe/H]=--0.67 (Z=0.006) and turnoff mass is about
0.852--0.868$~M_\odot$ (VandenBerg et al. 2006).

\section{Evolutionary model}

While calculating evolutionary models of binary stars, we must
take into account mass transfer and associated physical mechanism
which lead to mass and angular momentum loss. We use a formula
based on that used to calculate angular momentum loss via a
stellar wind (Paczy\'nski \& Zi\'o\l kowski 1967; Zi\'o\l kowski
1985 and De Greve 1993). We can express the change in the total
orbital angular momentum ($J$) of a binary system as

\begin{equation}
\frac{\dot{J}}{J} = f_{1} \: f_{2} \: \frac{M_{\rm 1} \:
{\dot{M}}_{\rm 2}}{M_{\rm 2} \: M_{\rm tot}},
\end{equation}

where, $M_1 $, $M_2 $ and $M_{tot} $ denote, respectively, the
mass of the primary, secondary and the total mass of the system,
$f_{1}$ is the ratio of the mass ejected by the wind to that
accreted by the primary component and $f_2$ is defined as the
effectiveness of angular momentum loss during mass transfer (Sarna
\& De Greve 1994, 1996).

Models of secondary stars filling their Roche lobes were computed
using a standard stellar evolution code based on the Henyey-type
code of Paczy\'nski (1970), which has been adapted to low-mass
stars (as described in detail in Marks \& Sarna 1998). We use the
Eggleton (1983) formula to calculate the size of the secondary's
Roche lobe.

For radiative transport, we use the opacity tables of Iglesias \&
Rogers (1996). Where the Iglesis \& Rogers (1996) tables are
incomplete, we have filled the gaps using the opacity tables of
Huebner et al. (1977). For temperatures lower than 6000 K, we use
the opacities given by Alexander \& Ferguson (1994) and Alexander
(private communication).

To understand the evolution of close binary V228 we computed
various evolutionary sequences: for different chemical
compositions Z=0.006--0.2; initial secondary masses
0.85--1.35$M_\odot $ and initial mass ratios ($q_i =
M_{1,i}/M_{2,i}$) from 0.6 to 0.95. For each system the secondary
fills Roche lobe with a small helium core (Hertzsprung gap).

The lower conservative limit for total mass of the system is about
1.7$~M_\odot$, which infer that the original primary had a mass
exceeding 0.85$~M_\odot $.

\section{Results}

From computed evolutionary sequences we predict that:

\noindent 1. Initial mass of the primary and secondary was about
0.85--0.9$~M_\odot $, and mass ratio around 1;

\noindent 2. Current properties of the system indicate that the
original primary filled Roche lobe in the Hertzsprung gap -- early
case B mass transfer;

\noindent 3. The best fitting model implies loss of about 50 per
cent of initial total orbital momentum, but only 5 per cent loss
of initial total mass;

\noindent 4. The initial parameters for the evolutionary model are
as follow: $M_{1,i} = 0.88~M_\odot $, $M_{2,i} = 0.85~M_\odot $,
$P_i=1.35 ~$days, $f_1$=0.05, $f_2$=4.65 and Z=0.006
([Fe/H]=0.67);

\noindent 5. The less massive component have a small helium core
of mass 0.12--0.17$~M_\odot $ and exchange mass in the nuclear
time scale;

\noindent 6. The best fitting model (0.88+0.85) spend about 10
Gyrs in detached configuration, while 0.2--0.3 Gyrs in
semidetached.


\acknowledgements 

This work were supported by grands 1 P03D 00128 and
76/E--60/SPB/MSN/P--03/DWM35/2005--2007 from the Ministry of
Sciences and Higher Education.


\end{document}